\documentclass[referee]{aa} 

\usepackage{graphicx}
\usepackage{color}
\usepackage[varg]{txfonts}   
\begin{document}

   \authorrunning{Li \& Peter}

   \title{Plasma injection into a solar coronal loop}

   \author{L. P. Li \inst{1,2,3} and
           H. Peter\inst{2}}

   \institute{CAS Key Laboratory of Solar Activity,
              National Astronomical Observatories, 
              Chinese Academy of Sciences, 
              100101 Beijing,
              People's Republic of China\\
              \email{lepingli@nao.cas.cn}
         \and
              Max Planck Institute for Solar System Research (MPS),
              37077 G\"ottingen, Germany
         \and
              University of Chinese Academy of Sciences, Beijing  100049, People's Republic of China
             }
   \date{Received ????; accepted ????}

  \abstract
   {
   The details of the spectral profiles of extreme UV emission lines from solar active regions contain key information to investigate the structure, dynamics, and energetics of the solar upper atmosphere.
   }
   {
   We characterize the line profiles not only through the Doppler shift and intensity of the bulk part of the profile. More importantly, we investigate the excess emission and asymmetries in the line wings   to study twisting motions and helicity. 
   }
   {
   We use a raster scan of the Interface Region Imaging Spectrograph (IRIS) in an active region. We concentrate on the \ion{Si}{iv} line at 1394\,\AA\ that forms just below 0.1\,MK and follow the plasma in a cool loop  moving from one footpoint to the other. We apply single-Gaussian fits to the line core, determine the excess emission in the red and blue wings, and derive the red-blue line asymmetry. 
   }
   {
   The blue wing excess at one footpoint shows injection of plasma into the loop that is then flowing to the other side. At the same footpoint, redshifts of the line core indicate that energy is deposited at around 0.1\,MK. The enhanced pressure would then push down the cool plasma and inject some plasma into the loop. In the middle part of the loop, the spectral tilts of the line profiles indicate the presence of a helical structure of the magnetic field, and the line wings are symmetrically enhanced. This is an indication that the loop is driven through the injection of helicity at the loop feet.  
   }
   {
   If the loop is driven to be helical, then one can expect that the magnetic field will be in a turbulent state, as it has been shown by existing MHD models.  The turbulent motions  could provide an explanation of the (symmetric) line wing enhancements which have been seen also in loops at coronal temperatures, but have not been understood so far.
   }


   \maketitle

\section{Introduction}\label{S:intro}

The spectral profiles of the extreme UV (EUV) emission lines from the Sun below $\approx$1600\,\AA\ contain essential information on the dynamic and thermal structure of the plasma in the solar upper atmosphere (e.g., Mariska \cite{mariska92}, Del Zanna \& Mason \cite{delzanna18}). It is quite common that optically thin emission lines in the EUV have non-Gaussian profiles with enhanced emission in the wings of  lines as first reported by Kjeldseth Moe \& Nicoles (\cite{moe77}). They pointed out that a single-Gaussian can fit the core of the transition region lines, but cannot account for the enhanced line wings. Dere \& Mason (\cite{dere93}) published a comprehensive summary of  normal and exotic transition region line profiles and confirmed that the profiles of \ion{C}{iv} and \ion{Si}{iv} are in general not well described by a single-Gaussian. Usually, the excess emission in the wings is interpreted as being due to non-resolved high-velocity flows. They could be caused by various  processes, such as twisting motions (De Pontieu \cite{depo14b}), field-aligned flows (De Pontieu \cite{depo09}), reconnection outflows (Dere et al. \cite{dere89}, Innes et al. \cite{innes15}), or even more exotic effects (Peter \cite{peter10}) such as non-Maxwellian velocity distribution functions.

In most cases where the transition region line profiles are not single-Gaussians, they are well described either by a double-  or a triple-Gaussian  profile. In the quiet Sun, two-component profile fits typically have a narrow core and a broad second component, which contributes up to 25\% to the total intensity above the chromospheric network (Peter \cite{peter00}). Peter (\cite{peter01}) proposed that the second component and the line core are formed in radically different physical regimes, namely small network loops and coronal funnels that are mixed along the line of sight. In active regions, Peter (\cite{peter10})  found that the spectra are mostly best fitted by a narrow line core and a broad minor component, too. An association of the minor component with propagating disturbances was then reported by Tian et al. (\cite{tian11b}). However, these two-component profiles might actually also be three-component profiles, with narrow components accounting to the excess emission in the two wings (Wang et al.~\cite{wang13}). However, mostly the spectral resolution is not really sufficient to distinguish these two cases (Peter \& Brkovi\'c \cite{peter03}). Only in cases like explosive events (e.g., Dere et al. \cite{dere89}) such a three-component structure is clear. More recent observations with the Interface Region Imaging Spectrograph (IRIS; De Pontieu et al. \cite{depo14a}) at unprecedented spatial and spectral resolution showed the existence of even more complicated spectra, e.g., with chromospheric absorption lines in the continuum and in strong transition region lines (Peter et al. \cite{peter14}), self-absorption in transition region lines (Yan et al.~\cite{yan15}), or complex bands of molecular lines in absorption and emission overlayed on the normal transition region spectra (Schmit et al.~\cite{schmit14}). All these effects are best seen in  UV bursts (Young et al.\ \cite{young18}).

Chromospheric, transition region, and coronal lines ubiquitously show line profile asymmetries. In chromospheric lines Ding \& Schleicher (\cite{ding98}) reported that the majority of \ion{Ca}{ii}\,K line profiles show a blue asymmetry. At loop footpoints De Pontieu et al. (\cite{depo09}) quantified the red-blue-asymmetry of line profiles and identified faint but ubiquitous upflows from the chromosphere. In the network regions, spectral lines show significant asymmetry in the blue wing of the emission line profiles, which was also considered as high-speed transition region and coronal upflows in the quiet Sun (McIntosh \& De Pontieu \cite{mcin09}). Using both models and simulations, Mart\'inez-Sykora et al. (\cite{mart11})  highlighted that the spectral asymmetry is sensitive to the velocity gradient with height (viz.,\ along the line of sight) in the transition region of coronal loops. In an active region, Brooks \& Warren (\cite{broo12}) identified regions of asymmetric profiles, and found that the red-blue-asymmetry is dependent on temperature. In some bright moss areas of two active regions, Tripathi \& Klimchuk (\cite{trip13}) measured the red-blue-asymmetry in line profiles formed over a wide range of temperatures, and derived emission measure distributions from the enhanced wing emission. They emphasized that the red-blue-asymmetry and associated emission measures are small.

Such non-Gaussians line profiles are also expected and found in hot active region structures. The nanoflare heating model predicts high-speed evaporative upflows (Antiochos \& Krall \cite{anti79}, Patsourakos \& Klimchuk \cite{pats06}, Li et al. \cite{li15}) that can show up in the spectra. Generating synthetic line profiles based on one-dimensional hydrodynamic simulations, Patsourakos \& Klimchuk (\cite{pats06}) predict distinctive enhancements in the blue wing of the \ion{Fe}{xvii} line profile. Such characteristics have been found in line profiles near the footpoints of active region loops (Hara et al. \cite{hara08}, Bryans et al. \cite{brya10}). Hara et al. (\cite{hara08}) observed significant deviations in the blue wing from a single-Gaussian profile and suggested that unresolved high-speed upflows occur there.  Tian et al. (\cite{tian11a}) reported that a faint excess emission at $\sim$100\,km/s in the blue wing of coronal emission lines generally accompanies the enhancement of the moments of the line profile, supporting the presence of unresolved upflows in active regions, too. Peter (\cite{peter10}) pointed out that in active region loops the excess emission in the line wing is seen simultaneously  in the blue and the red wings near the loop apex.This is unlikely to be caused by a field-aligned flow, because in that observation the line-of-sight seemed to be almost perpendicular to the loop. This raises the question if the explanation of the excess emission near loop footpoints in the the blue wing of the profile being due to upflows is unique, either.

The EUV\ observations mentioned above are based either on comparably large raster scans or sit-and-stare observations. Both are acquired with slit spectrographs and cannot disentangle the spatial and the temporal evolution. The raster maps provide only very limited time resolution (typically longer than many minutes). In the sit-and-stare mode the slit is at a fixed spatial position, so the field-of-view of spectroscopic information is limited to one dimension along the slit. Because loops are elementary components in the solar atmosphere covering a range of sizes and time scales  (e.g., Tian et al. \cite{tian09}, Peter et al. \cite{peter13}, Yan et al. \cite{yan13}), it is important to understand their dynamic and thermal characteristics. For example, using spectroscopic maps one can investigate the spatial structure of helicity in these loops (e.g., Li et al. \cite{li14}), but the question remains how these spatial structures in the spectroscopic data change in time.

In this paper, we investigate the temporal evolution of the spectroscopic features in a loop as they evolve in time. In the IRIS observations of the \ion{Si}{iv} line we present here, we have been lucky that the slit scanning the raster map was moving with just the right speed, so that it almost exactly followed a feature that was injected into the loop on one side and then propagated along the loop to the other footpoint. Because the apparent motion of the slit has the same velocity as the feature in the loop, we can follow the temporal evolution of the spectral features all along the loop.


\section{Observations and data processing}\label{S:obs}

\begin{figure}
   \includegraphics[width=88mm]{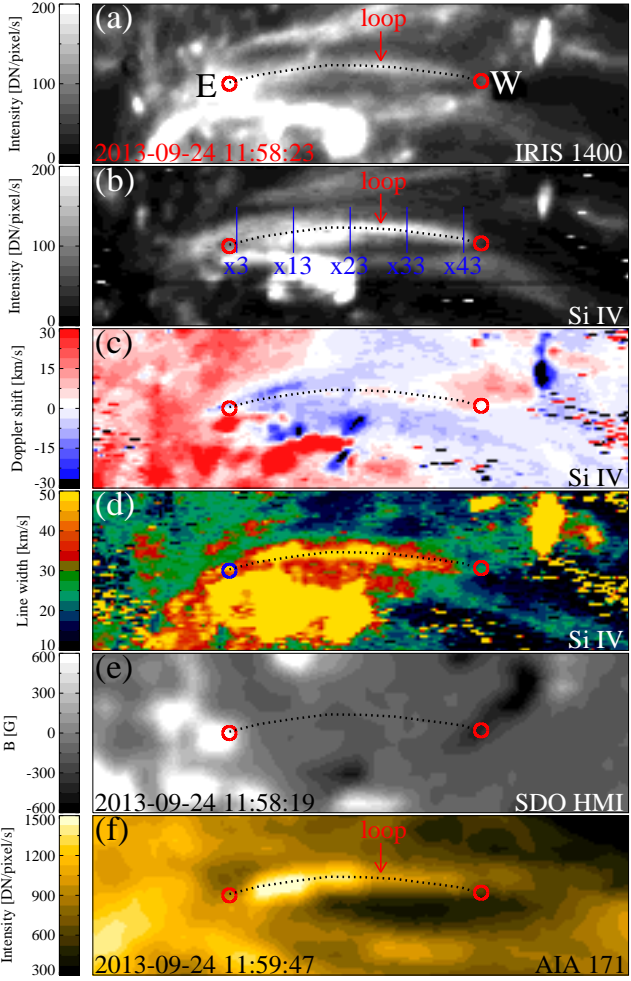}
   \caption{Cool loop and context from the IRIS and SDO data.
   Panel (a) shows a snapshot of the IRIS slit-jaw images in the 1400\,\AA\ channel,
   panels (b-d) maps of intensity (b), Doppler shift (c), and the  1/$e$ line
   width (d) of the \ion{Si}{iv} (1394\,\AA) line. 
    Panel (e) displays an HMI line-of-sight magnetogram and panel (f) a snapshot from the AIA images at 171\,\AA.
   The dotted lines indicate the \ion{Si}{iv} loop.
   The circles mark two footpoints of the loop with E and W indicating the eastern and western feet.
   The blue vertical lines x3, x13, x23, x33, and x43 in (b)
   indicate five samples of the spatial positions
   of the spectra showed in Fig.\,\ref{F:spectra}.
   The field-of-view is 33.2\,\arcsec$\times$10.0\,\arcsec \ with the center  at solar (X,Y) = (-251.9\,\arcsec, 56.4\,\arcsec).
See Sects.\,\ref{S:obs} and \ref{S:overall}.
    \label{F:loop}}%
\end{figure}

The IRIS observatory provides simultaneous spectra and images of the upper solar atmosphere. The active region investigated in this paper (NOAA AR 11850) was scanned several times by IRIS during its emergence in September 2013. In this active region, already hot explosions in the cool atmosphere of the Sun (Peter et al. \cite{peter14}), and conversion from mutual helicity to self-helicity (Li et al. \cite{li14}) have been reported. In this study we use the raster map from 11:44 UT to 12:04 UT on September 24, 2014. The loop we investigate here is located in the north of the active region (cf. Fig.\,\ref{F:loop}). We use  IRIS level\,2 data\footnote{Data are available at http://iris.lmsal.com.} which are already corrected for flat field, geometric distortion, and dark current.

A large dense raster was acquired with 400 raster steps of 0.35\arcsec\ each. The resulting field-of-view is of about 140\arcsec\ $\times$ 180\arcsec\ centered at about 82\arcsec\ north and 265\arcsec\ east of disk center. The exposure time of the individual spectra was 2\,s, the step cadence is 2.9\,s, and the spatial scale along the slit is about 0.17\arcsec\ per pixel. While all available wavelength bands have been recorded  in the slit-jaw images, here we use  only the 1400\,\AA\ channel. Taken together with the spectra, the 1400\,\AA\ images cover the corresponding field-of-view of the raster map, have a cadence of 12\,s, and a spatial sampling of about 0.17\arcsec\ per pixelö. Using the fiducial marks on the slit, we aligned all the IRIS data. For the spectroscopic analysis, we concentrate on the transition region line of \ion{Si}{iv} at 1394\,\AA, which is the stronger line of the \ion{Si}{iv} doublet accessible to IRIS. Further details of the observations can be found in the Supplementary Material S1 of Peter et al.~(\cite{peter14}).

For the Doppler maps in the \ion{Si}{iv} line a wavelength calibration is required. This is also provided with the level\,2 data, but here we checked and reiterated the calibration. For this, the average spectrum of the quiet plage area surrounding the emerging active region is used. In particular, for the  \ion{Si}{iv} line at 1394\,\AA\ we employ the nearby \ion{Ni}{ii} and \ion{Fe}{ii} lines originating in the chromosphere that should show small line shifts, below 1\,km/s (cf.\ Table\,\ref{table.lines}).  Considering also the uncertainties in the rest wavelengths we can expect the wavelength calibration to be limited to about 1\,km/s (see Supplementary Material S1 of Peter et al. \cite{peter14}), which is more than sufficient for the purpose of this study. In particular the average Doppler shift of the \ion{Si}{iv} line is about 6\,km/s to the red, consistent with previous observations (e.g., Peter \& Judge \cite{peter+judge99}).

\begin{table}
\caption{Lines of interest.}
\label{table.lines}
\centering 
\begin{tabular}{c c c c}
\hline\hline
line & rest wavelength [\AA] & $\Delta v$ [km/s] &
Doppler shift [km/s] \\
\hline
\ion{Fe}{ii} & 1392.816 & $-$203 & $-$0.5 \\
\ion{Ni}{ii} & 1393.33~\,  & $-$93  & $+$0.6 \\
\ion{Si}{iv} & 1393.76~\, & 0 & $+$5.9 \\
\hline
\end{tabular}
\tablefoot{
The lines of \ion{Fe}{ii} and \ion{Ni}{ii} are used for wavelength
calibration. The rest wavelengths are taken from Sandlin
et al. (\cite{sandlin86}), see also Supplementary material
S1 of Peter et al. (\cite{peter14}).
The column $\Delta v$ shows the difference
of the rest wavelengths to the \ion{Si}{iv} rest wavelength
in velocity units. The rightmost column lists the average
absolute Doppler shifts in a plage area with respect
to the respective rest wavelength. Positive values
correspond to redshifts. See Sect.\,\ref{S:obs}.}
\end{table}

To embed the IRIS observations into the context of the structure and evolution from the photosphere to the corona, we use data from the Solar Dynamics Observatory (SDO; Pesnell et al. \cite{pesn12}), namely of the Helioseismic and Magnetic Imager (HMI; Schou et al. \cite{scho12}) and the Atmospheric Imaging Assembly (AIA; Lemen et al. \cite{leme12}). The spatial and temporal sampling  of the HMI line-of-sight magnetograms are 0.5\arcsec\ per pixel and 45\,s. For the AIA multi-wavelength images the respective values are 0.6\arcsec\ per pixel and 12\,s. We spatially scaled the SDO observations to match the IRIS slit-jaw images at 1400\,\AA\ and aligned them using several characteristic features, such as sunspots, network, and plage patterns. The exact spatial alignment is not crucial for our study because the SDO data are mainly for context, and an accuracy of about 1\arcsec\ is sufficient (and easily achievable).

\section{Results}

\subsection{Overall properties of the emerging loop}\label{S:overall}

We  first investigate the general properties of the cool loop we see in the \ion{Si}{iv} line. To show the context of the observation we display the observed loop in its line profile parameters along with the underlying magnetic field and the coronal context in Fig.\,\ref{F:loop}. The overall comparison of the snapshot of the 1400\,\AA\ slit-jaw image (Fig.\,\ref{F:loop}a) and the raster map in the \ion{Si}{iv} intensity from the spectra (Fig.\,\ref{F:loop}b) shows that these are roughly similar. This demonstrates that in our observation in the loop the 1400\,\AA\ channel is indeed dominated by the transition region \ion{Si}{iv} line. However, some bright dots are visible in the 1400\,\AA\ channel not seen in the \ion{Si}{iv} line intensity, which are most likely due to the chromospheric contribution to the 1400\,\AA\ channel.

The most prominent feature in the field-of view in Fig.\,\ref{F:loop} is a cool loop visible in \ion{Si}{iv} that is stable at least during the 20\,min  when building up the raster map. The central  axis of the loop and its footpoints are highlighted in the figure (dotted line and red circles).
We identify the loop and its footpoints visually based on the IRIS slit-jaw images in the 1400\,{\AA} channel.
The corresponding HMI line-of-sight magnetogram (Fig.\,\ref{F:loop}e) shows, as expected, that the loop connects two opposite polarity plage-type features in the  active region. 
South of the loop is a highly dynamic region which hosts an UV burst which has been discussed by Peter et al. (\cite{peter14}) (their bomb B4 in their Fig.\,1). However, this UV burst is not connected to the loop we investigate here, at  least in the sense that the temporal variability of the loop is not related to the UV burst. This is clear by the movie provided with the study of Peter et al. (\cite{peter14}) (and also visible in the animation we include with Fig.\,\ref{F:coronal.comp}). Furthermore the spectra in the UV burst are radically different from the loop (cf. Fig. S6 of Peter et al. \cite{peter14}).    

For a first characterization of the \ion{Si}{iv} 1394\,\AA\  line profile we performed a single-Gaussian fit with a constant background for the continuum. Because we want to concentrate on the line core, we apply the fit only within $\pm$20\,km/s from the peak of the line (and the continuum).  To characterize the line core we use the  total intensity from the single-Gaussian fit, $T_{\rm{SG}}={\int}I_{\rm{SG}}\,{\rm{d}}v$ ($I_{\rm{SG}}$ is the intensity profile of the fit),  the Doppler shift, $v_{\rm{D}}$, and the Gaussian width, $w_{1/e}$ (half width at $1/e$ of the peak intensity).
Here we subtracted the instrumental width of about 26\,m{\AA} (De Pontieu et al. \cite{depo14a}) but not the thermal width. Because the  thermal width of \ion{Si}{iv} is small (less than 7\,km/s), the non-thermal broadening is very close to the line width (within 1\,km/s).
These line-core parameters are shown in Figs.\,\ref{F:loop}b--d.  The \ion{Si}{iv} intensity  shows the  loop throughout the raster, i.e., for at least 20\,min, which is much longer than the ionization and recombination time of \ion{Si}{iv} (cf. Peter et al. \cite{peter06}, their Fig.\,4). This indicates that the loop is stable and close to (or at least not too far from) ionization equilibrium. Therefore we can assume the loop to be a cool loop at a temperature close to the ionization equilibrium temperature of \ion{Si}{iv} at about 80\,000\,K (cf.\ Peter et al. \cite{peter06}, their Table 1 and Sect. 4.1). At least, there has to be a component of the loop (viz.,\ strand) relatively stable at this temperature.

The cool loop has a length of ${\approx}15.8$\arcsec\ (${\approx}12$\,Mm) and a width of ${\approx}0.8$\arcsec\ (${\approx}0.6$\,Mm). Here the width is measured as the full width at half maximum after subtraction of the background. For the width estimate we use the spatial range between lines x23 and x43 in Fig.\,\ref{F:loop}b, where the loop is most clearly defined with a background level of only about 20\%. This width roughly corresponds to the width of loops found in several-MK hot coronal loops (Aschwanden \& Boerner \cite{aschw11}) and is on the narrow side of the widths reported by Aschwanden \& Peter (\cite{aschw17}). In particular it also roughly matches the loop width measured in the observations of the first Hi-C rocket flight (Cirtain et al. \cite{cirtain13}) as seen in the 193\,\AA\ channel showing plasma at about 1.5\,MK. There Peter et al. (\cite{peter13}) found that these coronal loops have a smooth cross-sectional intensity profile with little (or no) indication of substructures, which one could interpret as the loops being spatially resolved. With IRIS and Hi-C having roughly the same spatial resolution (both about a factor of five better than AIA) this result suggests that there might be a common intrinsic width of cool loops (at 0.1\,MK) and hotter loops (at several MK) that deserves further investigation.

The large-scale dynamics in the cool loop are visualized by the Doppler map in Fig.\,\ref{F:loop}c. This  indicates that the eastern part (left) of the loop exhibits a blueshift, while the western part is redshifted. The red and blue pattern is consistent with the interpretation of a siphon flow along the cool loop (e.g., Bethge et al. \cite{beth12}). The blueshift in the background surrounding the loop in-between the main polarities of the emerging active region might be interpreted as being due to flux emergence between the main polarities of the active region. Such general uplifts of the upper atmosphere as a whole in response to flux emergence are also seen in the  simulation by Chen et al. (\cite{chen14}).

To show the internal dynamics within the loop, we display the line width in Fig.\,\ref{F:loop}d (Gaussian $1/e$ width). While in quiet Sun regions the (non-thermal line width) is of the order of 20\,km/s, at the location of the loop the line width is significantly larger (by a factor of about 2). Because the thermal width of \ion{Si}{iv} is only about 7\,km/s, the widths shown in Fig.\,\ref{F:loop}d are basically the same as the non-thermal line widths. Usually the non-thermal broadening is interpreted as being due to small-scale non-resolved (bulk and wave) motions in response to small-scale heating events (e.g.,\ Dere \& Mason \cite{dere93}, Chae et al. \cite{chae98}, Peter \cite{peter99}, \cite{peter01}). Thus one might attribute (at least part of) the non-thermal broadening to plasma heating along the loop. We emphasize that the width we find here is for the line core only, and that there is a significant contribution of the line wings to the total line intensity  (at times up to 40\%, see Sect.\,\ref{S:evolution.along.loop}). So to some small extent the large line width of the single-Gaussian fits in the loop will be also due to the strong wings excess.

Interestingly, the loop in \ion{Si}{iv} has a counterpart seen in the AIA channel at 171\,\AA\ (Fig.\,\ref{F:loop}f). In equilibrium the 171\,\AA\ channel is dominated by emission from \ion{Fe}{ix} forming at about 0.8\,MK, but it has also some minor contribution due to lines forming at lower temperatures around 0.25\,MK (cf. Boerner et al. \cite{boerner12}). So in principle it could be that the emission seen in the 171\,\AA\ channel is due to cool plasma. However, based on the greatly increased line width indicating strong heating in the cool loops (see Sect.\,\ref{S:disc.helical}) it is more probable that the loop seen in 171\,\AA\ is indeed due to hotter plasma of almost 1\,MK (see Sect.\,\ref{S:coronal.comp}). Therefore this loop might be an example for plasma at transition region temperature (of the order of 0.1\,MK) coexisting with coronal plasma at 1\,MK in the same loop. This would indicate that some strands of the loop are still at 0.1\,MK while other are already heated to coronal temperatures. Because the main emphasis of this study is on the non-Gaussian nature of the emission line profile in \ion{Si}{iv} we do not follow up this line of thought here.

\subsection{Propagation of features along the loop}\label{S:propagation}

\begin{figure}
   \centering
   \includegraphics[width=60mm]{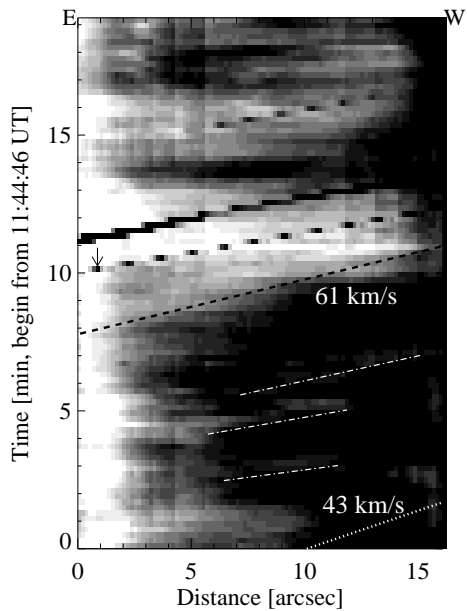}
   \caption{Proper motions along the loop.
   A space-time plot of a series of IRIS slit-jaw images
   in the 1400\,\AA\ channel along the dotted line EW as marked in Fig.\,\ref{F:loop}a.
   The dashed line and dash-dotted lines indicate proper motions
   along the loop. For comparison, the dotted line shows the sound speed near 80‡‡\,000\,K
   in the line formation region of \ion{Si}{iv}.
   The respective mean velocity is denoted by the numbers in the plot.
   The black arrow marks the IRIS slit scanning the loop.
   See Sect.\,\ref{S:propagation}.}
    \label{F:propermotion}%
\end{figure}

While overall the cool loop we study here is stable, numerous features can be seen that propagate along the loop, always in the same direction (east to west). To study this we investigate the slit-jaw images in the IRIS channel at 1400\,\AA\  that (in the loop) mainly shows the \ion{Si}{iv} line (cf.\ Sect.\,\ref{S:overall}). In particular, we check the temporal evolution of the 1400\,\AA\ emission along the dotted line in Fig.\,\ref{F:loop}a. The resulting space-time diagram is  shown it in Fig.\,\ref{F:propermotion}.

There are two prominent black linear features in Fig.\,\ref{F:propermotion}, one starting at the time 10\,min, and the other at 11\,min. The one at 10\,min with small spots (and marked by a small arrow) is the slit passing  across the loop. The thicker black line is due to a bad spot on the detector (masked in black in the slit-jaw images). Projected onto the Sun, these artifacts move with an apparent speed of about 80\,km/s.

The actual bright patches that move along the loop in east-west direction can be seen as inclined bright linear features, some of which are highlighted by the dotted lines. Typically these features move with roughly constant speeds ranging from some 45\,km/s  to 60\,km/s. This is slightly larger than the sound speed in the \ion{Si}{iv} source region of the emission of the cool loop (about 43\,km/s at 80\,000\,K). This implies that here we see disturbances propagating along the cool loop with about sound speed (e.g., waves or density enhancements). The individual bright features seen here have a spatial extent of  some 2\,Mm to 4\,Mm. The strong brightening feature in the loop appearing around the time when the slit scans the loop (just above the dashed line starting at 8\,min) is propagating with a speed of more than 60\,km/s.

By coincidence, the apparent motion of the slit along the loop has a speed similar to (though a bit faster than) the solar structures propagating within the loop. This is apparent when checking the propagating brightening moving along with the slit in Fig.\,\ref{F:propermotion}. This provides the nice possibility to follow the same structure as it moves and evolves along the loop. During the time the structure propagates along the loop (some two minutes) the difference between the apparent motion of the slit and the propagation of features along the loop translates into a difference in space of  only about 2\,Mm which is less than the size of the brightening. Also, from Fig.\,\ref{F:propermotion}. it is clear that the slit (black dots starting at ${\approx}10$\,min) are all the time well within the same bright feature moving along the loop. Thus in the following we will assume that the spectra taken while the slit scanned the loop do originate from the same feature that is injected at the eastern footpoint and then propagates westwards along the loop.

Before this most prominent propagating feature appears  (Fig.\,\ref{F:propermotion}, at time 8\,min), there are several smaller propagating brightenings starting from the eastern footpoint. While it would be very interesting to study the spectra of these precursors, too, spectroscopic data are available only for the propagating feature starting around time 10\,min.

\begin{figure*}
   \sidecaption
   \centering
   \includegraphics[width=120mm]{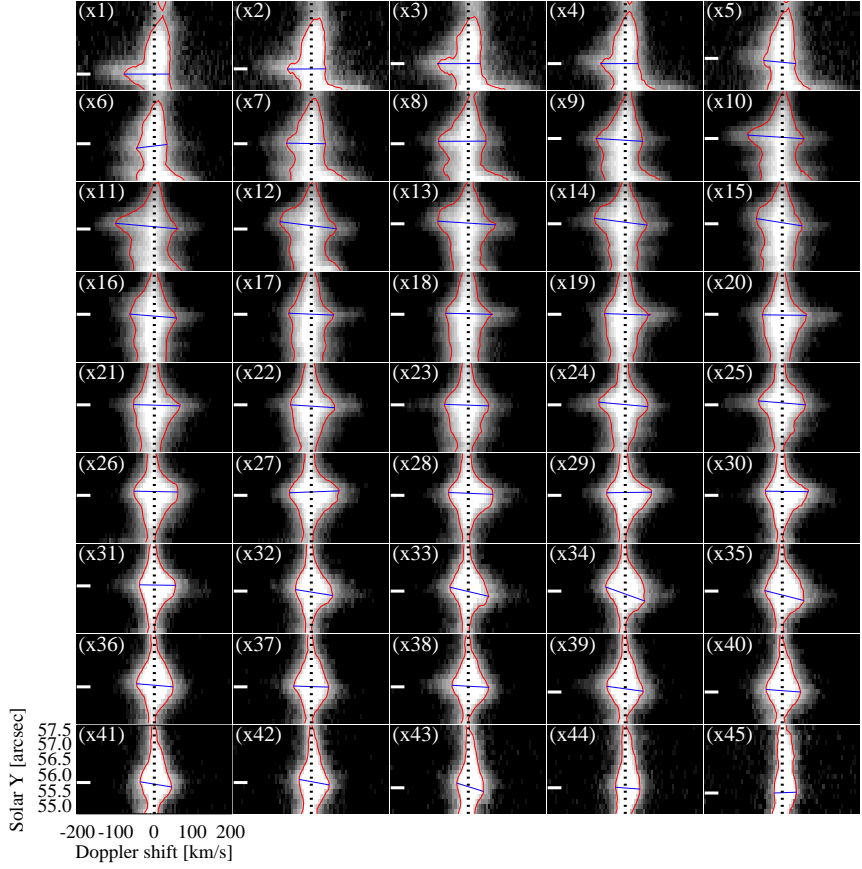}
   \caption{Spectral detector images of \ion{Si}{iv} (1394\,\AA) along the loop.
The 45 raster steps from the eastern to the western footpoint in Fig.\,\ref{F:loop} are labeled x1 through x45. The locations of x3, x13, \dots, and x43 are marked in Fig.\,\ref{F:loop}b
   by five blue vertical lines.
   The white markers indicate the location of
   the central axis of the loop as displayed by the dotted line in Fig.\,\ref{F:loop}b.
   The vertical dotted lines indicate zero Doppler shift.
   The red curves show the contours of the respective spectra at a level of 18\% of
   its maximum.
   The blue lines connect two positions with
   the maximum enhancements in the blue and red wings. See Sects.\,\ref{S:spectra} and \ref{S:quanti.detector.images}.}
    \label{F:spectra}%
\end{figure*}

\subsection{Non-Gaussian line profiles in the loop: injection and wing enhancement}\label{S:spectra}

The Doppler shifts derived from the  core of the \ion{Si}{iv} profile indicate a smooth siphon flow from the east to the west side (Sect.\,\ref{S:overall}, Fig.\,\ref{F:loop}c). In contrast, the emission in the wings of the \ion{Si}{iv} line exhibit a complex evolution while the plasma evolves from the injection at the eastern footpoint and then propagates to the west along the loop.

As a first step to investigate the \ion{Si}{iv} line profile along the loop we display the corresponding parts of the detector images in Fig.\,\ref{F:spectra}. We use the following housekeeping  for the spectra: the 45 raster positions  from the eastern to the western footpoint along the solar-x direction are labeled x1{\dots}x45. As a reference, in the context figure (Fig.\,\ref{F:loop}b) we indicate these positions for x3, x13, x23, x33, and x43 by blue vertical lines. The length of each of these lines  is identical to the spatial coverage (along the slit) of the detector images in Fig.\,\ref{F:spectra}. The white marker in each of the spectra in Fig.\,\ref{F:spectra} indicates the central axis of the loop, i.e., the location of maximum brightness across the loop (which is the same as the
dotted line in Fig.\,\ref{F:loop}b).

To highlight the asymmetry of the spectra, we also show contour lines of the spectral detector images in Fig.\,\ref{F:spectra}. Here we show contour levels of 18\%, i.e., at each spatial position we mark the wavelength (or Doppler shift) at which the intensity dropped to 18\% of the peak intensity at that location. This choice of 18\% is motivated by the residual of the Gaussian fits, which on average is 18\% in the red and 21\% in the blue wing. We simply took the smaller number. We also checked the contour levels at 15\%, 20\%, 25\%, and 30\%, but the results are qualitatively the same.

\begin{figure}
   \centering
   \includegraphics[width=88mm]{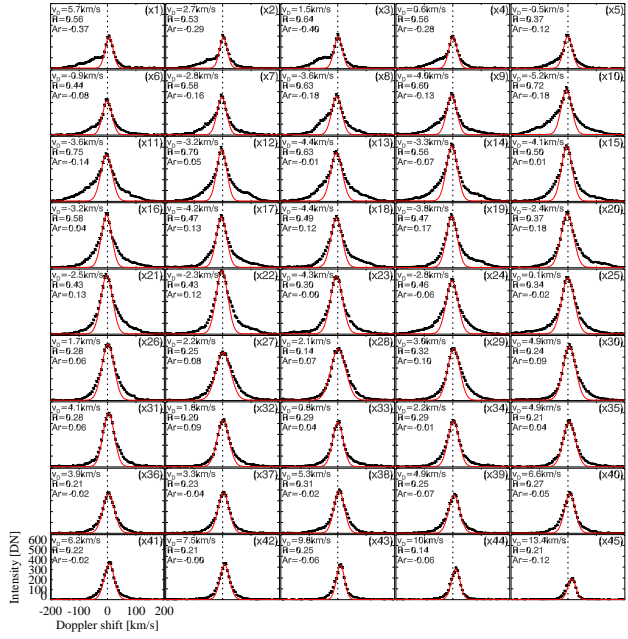}
   \caption{Individual spectral profiles of \ion{Si}{iv}
(1394\,\AA) along the loop.
   Each profile was recorded in one spatial pixel
on the central axis of the loop (i.e.,\ maximum emission from the loop at the respective slot position; dotted line in Fig.\,\ref{F:loop}b). These are the spectral profiles at the locations of the short horizontal white markers in the detector images from Fig.\,\ref{F:spectra}. 
   The diamonds (with the bars) show the observed spectra
   and the red lines indicate single-Gaussian fits
to the line core. To ensure that the single-Gaussian
fits represent the line core only, the fit was performed only within $\pm$20\,km/s
 from the peak of the line.
   The parameters denoted by the numbers in the plots
show line centroid
   ($v_{D}$) of the red fits, the total residual of
the intensity ($R$),
   and the red-blue asymmetry ($A_{\rm{r}}$). See Sects.\,\ref{S:spectra} and \ref{S:evolution.along.loop}.}
    \label{F:lineprofiles}%
\end{figure}

Near the eastern footpoint (x1) the detector images and the contours show a very clear signal of a one-sided excess emission in the blue wing, indicating the injection of plasma through an upflow into the loop (Fig.\,\ref{F:spectra}-x1). Away from the footpoint, from position x2 to x5, the clear single-sided excess in the blue wing ceases. At these positions no clear excess emission is seen in the red wing. This is also confirmed by the plots of the spectral profiles on the central axis of the loop in Fig.\,\ref{F:lineprofiles}. From east to west,  Fig.\,\ref{F:lineprofiles} shows the spectral profiles at the respective location of the peak intensity across the loop (i.e., at the location of the white markers in Fig.\,\ref{F:spectra}). Here the comparison to the single-Gaussian fits clearly reveals the line asymmetry, and at locations x1 to x5 the clear single-sided enhancements in the blue wing. Starting with position x6 there is significant excess emission in both wings, with a peak around x11. Finally towards the western footpoint (x45) both the detector images in Fig.\,\ref{F:spectra} and the line profiles in Fig.\,\ref{F:lineprofiles} show a decrease of the excess emission in the wings. Actually, near the western footpoint the profiles are close to single-Gaussians.

This simple check of the spectra already reveals the pattern of injection at one footpoint and enhanced emission in both wings through the majority of the loop before turning into a more single-Gaussian profile at the other footpoint. With the discussion from Sect.\,\ref{S:propagation} this implies that while a parcel of plasma flows from east to west it changes its spectroscopic properties over time after the injection from non-Gaussian to close to single-Gaussian. In the following we will quantify this impression by first characterising the spectral tilt in the detector images (Sect.\,\ref{S:quanti.detector.images}) and then investigating the line asymmetries (Sect.\,\ref{S:evolution.along.loop}).

To quantify the non-single-Gaussian nature of the line profiles several options are available. For example, one could fit a kappa distribution to the spectra (e.g.,\ Dudik et al. \cite{dudik17}). This could account for the enhanced emission in the line wing, but not for the asymmetries we see. This is why we do not consider this option. Another possibility would be to use double-Gaussian fits (e.g., Peter~\cite{peter10}). There a second broad Gaussian component would account for the excess emission in the line wing, and the relative shift of the core and the wing component would account for the line asymmetry. In this scenario one implicitly assumes that there is exactly one additional flow component in the loop that would account for the line wing excess. Firstly, one could then wonder why not to use even a three-component fit (e.g., Wang et al.~\cite{wang13}). Secondly, and more importantly, should turbulence evolve in the loop (as discussed in Sect.\,\ref{S:disc.helical}) such a representation would be misleading. In that case there would not be a distinct second flow component in the loop, but the excess emission in the line wing would be caused by the (continuous) distribution of the velocity within the turbulent medium. Yet another possibility to characterise the non-Gaussian profiles is to quantify the excess emission in the line wings (with respect to a single Gaussian) and the line profile asymmetry. This procedure does not use a priory assumption on the physical process causing the non-Gaussian nature of the profile and it allows a quantitative comparison to future models for the line profiles in loops. This is why we follow this latter option in Sect.\,\ref{S:evolution.along.loop}.

\subsection{Quantifying the spectral tilt and helical flow}\label{S:quanti.detector.images}

If a line profile gradually shifts its position (or centroid) with the spatial direction along the slit, the resulting detector image of the line profile will be tilted. This can be illustrated by the detector image at position x33 along the loop as shown in Fig.\,\ref{F:spectra}: With increasing position of solar\,Y along the slit (from about 56.0\arcsec\ to 56.5\arcsec) the line shift seems to move gradually from the red to the blue. Such a change of line shift with position across the loop  can indicate a rotational component of the flow in the loop. Here at smaller solar\,Y values the plasma moves away from the observer, at larger solar\,Y it moves towards the observer. Together with the flow along the loop this would indicate a helical flow along the loop (e.g., Li et al. \cite{li14}).

To quantify the spectral tilt, we use the red contour lines (at the 18\% level\footnote{The actual results do not depend on the exact choice of the level. See also Sect.\,\ref{S:spectra} for a discussion on choosing the 18\% level.}) of the detector images in Fig.\,\ref{F:spectra}. We determine the maximum blueshifts ${\Delta}v_{\rm{blue}}$ and redshifts ${\Delta}v_{\rm{red}}$ (both numbers defined to be positive) of these contours with respect to the Doppler shift $v_{\rm{D}}$ of the line core (cf.\ Sect.\,\ref{S:overall}) at the respective location. The Doppler dispersion (viz., width of the 18\% contour level) is then given by the sum of these two, ${\Delta}v={\Delta}v_{\rm{red}}+{\Delta}v_{\rm{blue}}$. Recording the position along the slit where the maximum redshift and blueshift occur, we can then determine the spatial offsets ${\Delta}y$ between the maximum redshift and blueshift of the contours. This yields the spectral tilt, ${\Delta}y/{\Delta}v$. Essentially, the spectral tilt is a measure for the slope of the line connecting the location of maximum blueshift and redshift in each detector image (i.e., the blue lines in Fig.\,\ref{F:spectra}; e.g., see x33 for a good example). If there is no evident maximum red or blue wing contour (e.g., with x1 in Fig.\,\ref{F:spectra}), we assume that ${\Delta}y=0$. Of course, the  values for the quantities ${\Delta}v_{\rm{blue}}$, ${\Delta}v_{\rm{red}}$, ${\Delta}y$, and ${\Delta}y/{\Delta}v$ will depend (slightly) on the choice of the contour level (here 18\%). However, we use ${\Delta}y/{\Delta}v$ mainly to investigate systematics of the spectral tilt and as such, the actual values are not important.

\begin{figure*}
   \centering
   \includegraphics[width=180mm]{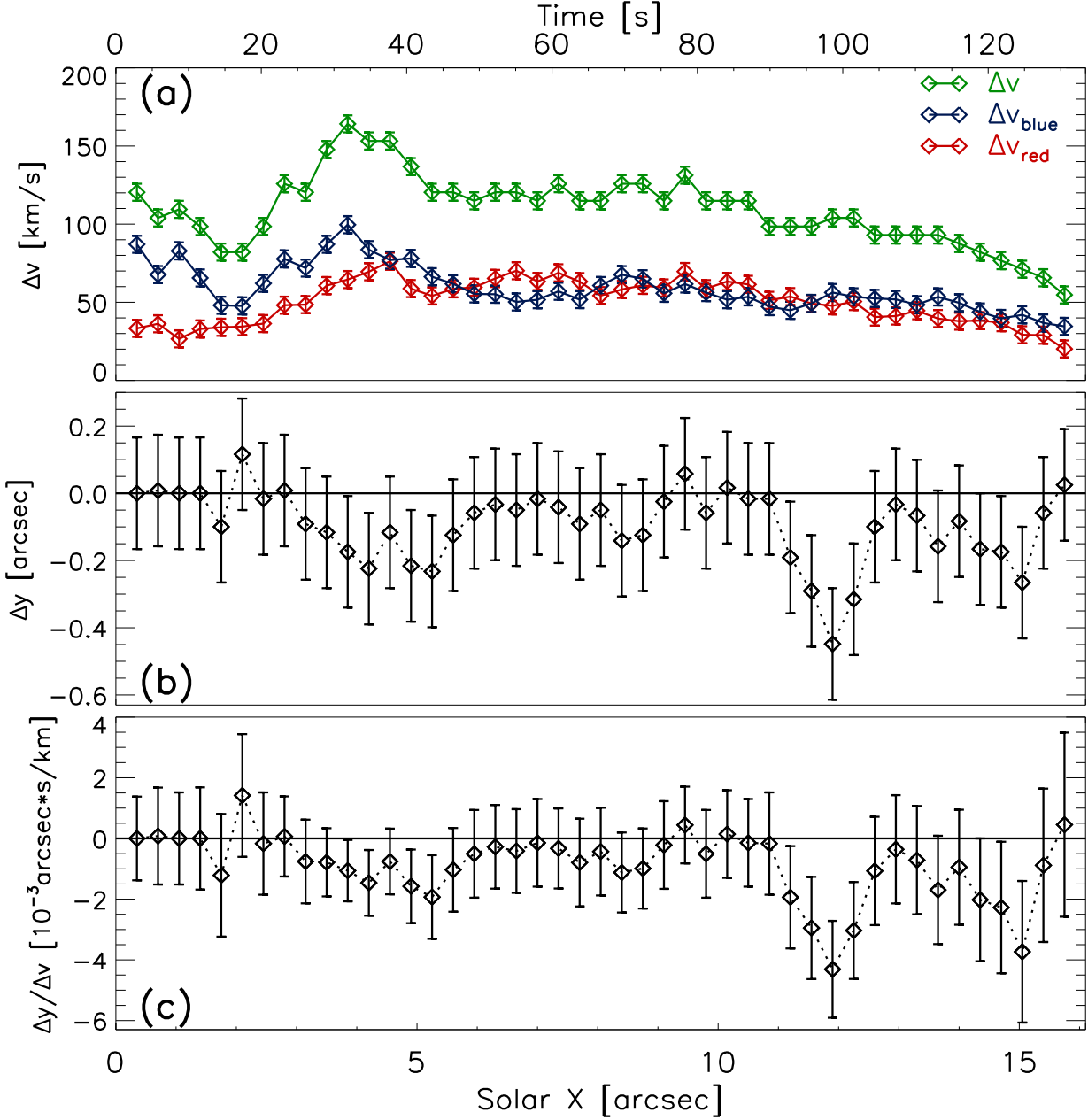}
   \caption{Evolution of spectral tilt of \ion{Si}{iv} (1394 \AA) along the loop. 
   Panel (a) shows the Doppler dispersion separately
for the blue wing (${\Delta}v_{\rm{blue}}$), the red
wing (${\Delta}v_{\rm{red}}$), and its combination
(${\Delta}v$).
   Panel (b) displays the offset between the maximum
enhancements in the blue and red wings along the slit
(${\Delta}y$). The errors are conservatively estimated
to be $\pm1$ spatial pixel.
   Panel (c) shows the spectral tilt (${\Delta}y/{\Delta}v$).
   Because we follow a parcel of gas moving along the loop, the top axis shows the time to indicate the temporal evolution. See Sect.\,\ref{S:quanti.detector.images}.}
    \label{F:spectraltilts}%
\end{figure*}

The Doppler dispersions ${(\Delta}v$, ${\Delta}v_{\rm{red}}$, ${\Delta}v_{\rm{blue}}$), the spatial offset (${\Delta}y$), and the spectral tilt (${\Delta}y/{\Delta}v$) are shown along the loop in Fig.\,\ref{F:spectraltilts}. The Doppler dispersion peaks around 3\arcsec\ to 4\arcsec\ away from the eastern footpoint (just as the line asymmetry does; see Sect.\,\ref{S:evolution.along.loop}). However, no clear trend can be seen in the spatial offset (as a very rough error estimate for  ${\Delta}y$ we conservatively  assume $\pm1$ spatial pixel). Also the spectral tilt ${\Delta}y/{\Delta}v$ does not show a clear trend. However, one systematic is prominent: While there are not too many locations with a significant spectral tilt, the sign of the tilt is always the same. This implies that our results would be compatible with a systematic rotational component of the flow and thus consistent with a helical flow along the loop.
%
This spectral tilt is caused almost only by the excess emission in the line wings. The position of the line core (as characterised by the single Gaussian fit) does not show a variation across the loop, viz.\ along the slit. This is evident by inspection of Fig. 1c, where no significant systematic change of the line shift is visible perpendicular to the dotted line indicating the loop position.

Because the slit follows the flow along the loop our finding does not necessarily imply a helical flow all along the loop all the time. We can only state that we see a systematic rotational component of the flow as it moves along the loop. Hence we include a time axis at the top of Fig.\,\ref{F:spectraltilts}.

\subsection{Quantifying the evolution of the plasma parcel moving along the loop}\label{S:evolution.along.loop}

To  quantify the evolution of the plasma parcel moving along the loop, we characterize the line core parameters and the asymmetry and excess emission in the wings of the \ion{Si}{iv} profiles. 
This analysis uses the single-Gaussian fits of the line core (cf.\ Sect.\,\ref{S:overall}) which provide the total intensity of the line core, $T_{\rm{SG}}={\int}I_{\rm{SG}}\,{\rm{d}}v$, and 
its Doppler shift $v_{\rm{D}}$.
We use $I_{\rm{SG}}$ and $I_{\rm{obs}}$ to denote the profile of the single-Gaussian fit (of the core only) and the actually observed profile.
Then the residual of the intensity in the blue wing
can be defined as
\begin{equation}\label{eq:Rblue}
R_{\rm{blue}}=\frac{1}{T_{SG}} \int_{-\infty}^{v_{\rm{D}}} (I_{\rm{obs}}-I_{\rm{SG}})\,{\rm{d}}v .
\end{equation}
Likewise, the residual in the red wing is
\begin{equation}\label{eq:Rred}
R_{\rm{red}}=\frac{1}{T_{SG}} \int_{v_{\rm{D}}}^{-\infty}
(I_{\rm{obs}}-I_{\rm{SG}})\,{\rm{d}}v .
\end{equation}
We can now define the total residual as
\begin{equation}\label{eq:R}
R = R_{\rm{red}} + R_{\rm{blue}}
\end{equation}
and the red-blue asymmetry as
\begin{equation}\label{eq:Ar}
A_{\rm{r}}= R_{\rm{red}} - R_{\rm{blue}} .
\end{equation}

\begin{figure}
   \centering
   \includegraphics[width=88mm]{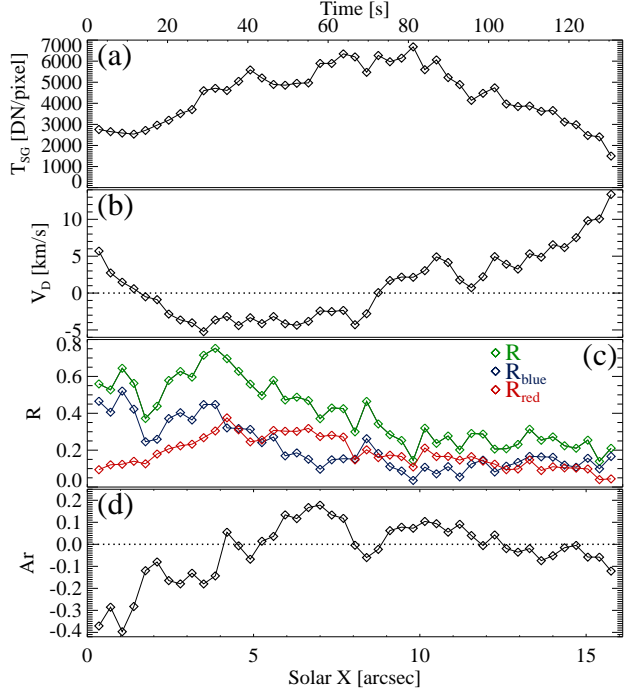}
   \caption{Evolution of the line profile characteristics of \ion{Si}{iv} (1394 \AA) along the loop.
   Panels (a) and (b) show the total intensity ($T_{\rm{SG}}$) and Doppler shift ($v_{\rm{D}}$) of the
single-Gaussian fits
   to the line core.
   Panel (c) displays the residuals of the intensity in
   the blue wing ($R_{\rm{blue}}$), the
red wing ($R_{\rm{red}}$), and over the whole line ($R$) as defined in Eqs.\,(\ref{eq:Rblue}) to (\ref{eq:R}).
   Panel (d) shows the
   red-blue asymmetry ($A_{\rm{r}}$) as defined in Eq.\,(\ref{eq:Ar}).
Because we follow a parcel of gas moving along the
loop, the top axis shows the time to indicate the temporal
evolution. See Sect.\,\ref{S:evolution.along.loop}.
   }
    \label{F:statistics}%
\end{figure}

The residuals provide the information on how strongly the observed profile deviates from the Gaussian fitting the core --- for the blue and red wing separately and for the whole profile. These are relative measures, and we find that the excess emission in both the blue and the red wings can reach up to 40\% of the total intensity of the Gaussian fit to the line core (e.g., in Fig.\,\ref{F:statistics}c at about 4\arcsec\ from the eastern footpoint we find ${R_{\rm{blue}}}{\approx}0.5$, ${R_{\rm{red}}}{\approx}{0.4}$). The total residual $R$ reaches values of up to 0.8, i.e.,\ the excess emission in the wings can be almost as strong as the line core emission.

The technique of red-blue asymmetry as used here, was introduced (in a different form) by De Pontieu et al. (\cite{depo09}). For their analysis they used  two (narrow) wavelength regions (or velocity offsets) symmetrically arranged with respect to the centroid  of
the profile (see also McIntosh \& De Pontieu \cite{mcin09}, Tian et al.
\cite{tian11a}). In our study, we consider the whole wavelength range (formally from $\pm\infty$ to $v_{\rm{D}}$). Qualitatively the results are the same as when having a more restricted wavelength (or velocity) range.
Most importantly, we do not only consider the red-blue asymmetry, $A_{\rm{r}}$, as done before, but also the residual $R$. This is important for the following simple reason. Even if there is no asymmetry, $A_{\rm{r}}{\approx}0$, there might be a significant deviation from a (single) Gaussian profile as emphasized already by Peter (\cite{peter10}). In our data set this is the case, e.g., around 5\arcsec\ from the eastern footpoint (Figs.\,\ref{F:statistics}c,d): here both the residuals in the red and blue wing are about 0.3, and there is no asymmetry (cf.\ profile x13 in Fig.\,\ref{F:lineprofiles}). Consequently, one should not look at the red-blue asymmetry alone, but also investigate the excess emission in the line wings, here characterized through $R$, $R_{\rm{red}}$, and $R_{\rm{blue}}$.

The variation of the line profile parameters indicate not only a spatial variation, but mainly a temporal variation of a plasma parcel moving along the loop (cf.\ Sect. 3.1). To illustrate this variation, we plot in Fig.\,\ref{F:statistics} the (line core) intensity, $T_{\rm{SG}}$, the Doppler shift, $v_{\rm{D}}$, the residuals, $R$, and the line asymmetry, $A_{\rm{r}}$, along the loop. The total intensity, $T_{\rm{SG }}$ (Fig.\,\ref{F:statistics}a), peaks when the plasma parcel reaches the middle (apex) of the loop. During the injection and also when it reaches the opposite footpoint the  intensity is lower.
%
Checking the density-sensitive lines of \ion{O}{iv} around 1400\,{\AA} we find that there is only a small change (less that 10\%) of the ratios from the loop footpoint to the apex. This applies to both ratios, 1401/1400 and 1401/1404. Thus we would expect only a modest change of density from footpoint to apex that cannot account for the intensity increase. Instead we argue that a small temperature change is responsible for the intensity variation (see Sect.\,\ref{S:disc.helical}).
%
The line centroid of the line core profile, $v_{\rm{D}}$ (Fig.\,\ref{F:statistics}b), shows a redshift in the immediate vicinity of the eastern footpoint (closer than 1.5\arcsec), which is also where the strongest excess emission in the blue wing is found (see below). In the major eastern part the shift is then towards the blue, and in the middle of the loop it turns to red. This is consistent with a continuous flow from east to west along the loop, where on the eastern half (``left'' of the loop apex) we see the projected upflow and in the western half we see the downflow. These projected flows speeds near the footpoints are of the order of 5\,km/s (cf.\ Fig.\,\ref{F:statistics}b).

The enhancement in the blue wing, $R_{\rm{blue}}$, is highest at the eastern footpoint consistent with an injection of plasma (Fig.\,\ref{F:statistics}c), because here the excess emission in the red wing is rather small. Inspection of the profiles (x1 to x3 in Fig.\,\ref{F:lineprofiles}) indicates that this upflow reaches speeds of 50\,km/s to 100\,km/s.  So within 1\arcsec\ to 1.5\arcsec\ from the eastern footpoint we see a high-speed upflow into the loop, while the bulk part of the plasma imaged here by the \ion{Si}{iv} line shows a net downflow ($v_{\rm{D}}$ being about 5\,km/s to the red). The high-speed injection is strong, because its $R_{\rm{blue}}$ is up to 0.5, implying that the upflowing plasma emits 50\% of the \ion{Si}{iv} emission compared to the source region of the line core associated with the slow downflow.

The excess emission in the red wing behaves quite different from the blue wing. While the blue wing drops more or less continuously from the eastern to the western side of the loop (with some fluctuations), the excess in the red wing, $R_{\rm{red}}$, first increases within the first 5\arcsec\ from the eastern footpoint. When it reaches its maximum, the excess in the red wing roughly matches that in the blue wing, $R_{\rm{red}}{\approx}R_{\rm{blue}}$. From there on to the western footpoint the excess emission in the red and blue wings very roughly match and more or less monotonically drop (even though there is some fluctuation). This implies that at the western footpoint the line profile is close to a (single) Gaussian. This quantitative result matches the visual impression of the line profiles in Fig.\,\ref{F:lineprofiles}.

The evolution of the excess emission in the blue and red wing, of course, leaves its mark in the red-blue asymmetry of the profile, $A_{\rm{r}}$ (Fig.\,\ref{F:statistics}d). This asymmetry is very strong only close to the footpoint, while over most part of the loop (more than 4\arcsec\ away from the eastern footpoint) the  asymmetry is comparably small. Clearly, with only looking at the red-blue asymmetry, one would not recognise that there is a significant excess emission in both line wings with roughly the same magnitude that is decreasing while the gas parcel moves along the loop to the footpoint opposite of where it was injected.

While the above results have been derived only for the spine of the loop (viz.,\ the central axis of the loop, see dotted line in Fig.\,\ref{F:loop}b), they are not special for this thin on-axis part of the loop. To demonstrate this, in Appendix\,\ref{appendix} we show the line profiles two IRIS pixels (or 0.33\arcsec) to the north and south of the central axis of the loop (Figs.\,\ref{lineprofiles_m} and \ref{lineprofiles_p}), i.e., at the edge of the loop. Likewise, we repeat plots with the variation of the various line profile parameters along the loop, once for the locations two pixels to the north and south (Fig.\,\ref{statistics_mp}) and for averaging the profiles in the north-south direction (Fig.\,\ref{statistics_av}). This makes clear that the results we showed in this paper are robust and reliable, and do not depend on the exact spatial positions we chose within or across the loop.

\section{Discussion}\label{S:disc}

The IRIS slit scanned the \ion{Si}{iv} loop with roughly the same speed as the plasma within the loop is moving from east to west. This lucky coincidence provides a nice opportunity to follow the same gas parcel as it moves along the loop and study its evolution.

\subsection{Loop geometry}\label{S:disc.geometry}

The cool loop seen here is (most probably) a rather flat loop. The proper motion along in the loop from east to west is of the order of 60\,km/s. The blueshift on the eastern and the redshift on the western side are about 5\,km/s. Considering a simple geometry, the inclination angle of the flow, and thus the angle of the magnetic field channeling the flow, near its footpoints is only of the order of 5$^\circ$. Considering that the footpoint distance of the loop is just below 16\arcsec\ (or 12\,Mm), the apex height of the loop should be below 1\,Mm.\footnote{Actually, if the loop would be a segment of a circle with an angle of 5$^\circ$ and a chord length of 12\,Mm, its apex height would be 0.25\,Mm.} So this cool loop would be really flat. For comparison, the pressure scale height at the line formation temperature of \ion{Si}{iv} ($\approx$80\,kK) is about 5\,Mm. Therefore, the cool loop extends in height only over a small fraction of the scale height and hence the density in the loop should not change significantly (through gravitational stratification). Thus for a constant mass flux we would also expect a roughly constant (proper) motion along the flat loop, which is what we see (cf.\ dash-dotted lines in Fig.\,\ref{F:propermotion}).

\subsection{Heating and injection of plasma at loop footpoint}\label{S:disc.heating}

Transiently brightening features at the eastern footpoint expand into the loop and we then see the plasma moving along the loop (Sect.\,\ref{S:propagation}). Together with the clear excess emission in the blue wing (Sects.\,\ref{S:spectra} and \ref{S:evolution.along.loop}) at the eastern footpoint, this strongly supports that there is an injection of plasma. Such injections at loop footpoints have been suggested before to supply mass to coronal structures (Xia et al. \cite{xia03}, De Pontieu et al. \cite{depo09}, Guo et al. \cite{guo10}, He et al. \cite{he10}, Tian et al. \cite{tian11b}), because these would naturally cause enhancements in the blue wing (De Pontieu et al. \cite{depo09}, McIntosh \& De Pontieu \cite{mcin09}, Tian et al. \cite{tian11a}, Brooks \& Warren \cite{broo12}). Alternatively, the excess emission we see in the blue wing might be also due to high-speed evaporative upflows as predicted in the nanoflare heating model (Patsourakos \& Klimchuk \cite{pats06}). However, those models would predict the upflows at very high temperatures (several MK) (Li et al. \cite{li15}), while here we see the spectral signature in \ion{Si}{iv} that shows plasma below 0.1\,MK.

The spectral properties of the \ion{Si}{iv} profile show an interesting feature, that seems to contradict itself at first sight. As mentioned above, the excess emission in the blue wing as well as the proper motions indicate that plasma is injected into the loop. However, the core of the line profile shows a clear net redshift near the eastern footpoint.
%
As expected from the general transition region redshifts, at this loop footpoint also \ion{C}{ii} at 1336\,{\AA} and \ion{O}{vi} at 1401\,{\AA} show a net redshift. The redshift of \ion{O}{iv} is 2\,km/s to 3\,km/s larger than the redshift of the \ion{Si}{iv} core. \ion{C}{ii} is not straight forward to interpret because of its self-absorption feature.
%
Actually, the line core redshift and the strong line asymmetry to the blue side seem both to be restricted to the first 2\arcsec\ from the footpoint (cf.\ Figs.\,\ref{F:statistics}b and d). This suggests a fast up- and slow downflow to co-exist in the same resolution element of the observation.

This can be understood in the following scenario. If there is strong heating (with high energy input per particle) at around or just above the formation temperature of \ion{Si}{iv}, say at a few 0.1 MK, then at that location the pressure would be enhanced locally. This is consistent with the findings in 3D MHD models that predict a maximum of the energy input per particle at temperatures between 0.1\,MK and 1\,MK (Hansteen et al.~\cite{hansteen10}, Bingert \& Peter~\cite{bingert11}).
In response, the bulk of the plasma below would be pushed down causing a redshift of the line core. Because this pushes the plasma into the higher-density regions below, the speed would be rather limited. The enhanced pressure would then also accelerate some of the plasma upwards and because it moves into the thinner (hotter) parts, this upwards acceleration would cause higher speeds, just as observed: the excess emission in the blue wing is found around 50\,km/s or even more, while the core of the profile is shifted only by a few km/s (cf.\ x1 to x3 in Fig.\,\ref{F:lineprofiles}). This is similar to the numerical models by Spadaro et al.\ (\cite{spadaro06}) and  Hansteen et al.\ (\cite{hansteen10}). They see cool downward moving plasma and upward accelerated plasma that is heated (and seen then at higher temperatures) but accelerated by a much smaller degree (blueshifts of only a few km/s). In our observation we see a coronal signature (i.e.,\ heated plasma) in the loop, too, which is clearly visible in the 171\,\AA\ observations (see Fig.\,\ref{F:loop} and Sect.\,\ref{S:coronal.comp}). However, there is also a cool loop at the same location, and the injected plasma (seen in the line wing) reaches much higher speeds on the order of 50\,km/s. In conclusion, this observation of the downward motion seen in the core of the line profile while at the same time there is an excess emission and strong asymmetry in the blue wing of the same \ion{Si}{iv} line poses a challenge for current models.

\subsection{Helical motion and turbulence along the loop}\label{S:disc.helical}

In our observations we find two spectral features that might be tightly connected: the presence of a spectral tilt  (Sect.\,\ref{S:quanti.detector.images}) and the (almost) symmetric excess emission in the line wings (Sect.\,\ref{S:evolution.along.loop}) in the middle part of the loop, i.e.,\ away from the loop footpoints. These two features might be related to the presence of a helical flow and to the turbulent state of the magnetic field in the loop.

Spectral tilts can be interpreted by a helical motion in the loop. The spatial offset of red and blue shifts on the two sides of the loop (cf. Fig.\,\ref{F:spectra} and Sect.\,\ref{S:spectra}) indicate the presence of a rotational component of the flow. Together with the observed (proper) motion along the loop this indicates a helical motion of the plasma. At the transition region temperatures where \ion{Si}{iv} forms,  the atmosphere should be in a low plasma-$\beta$ stage (Peter et al.\ \cite{peter06}; their Fig.\,12a). Consequently, the flows should be parallel to the magnetic field and the helical flow of the plasma indicates that the magnetic field in the loop is helical, too. Such helical (or twisting) motions have been seen before. For example, Li et al.\ (\cite{li14}) interpreted the changes of the spectral tilt in time as being due to the change from mutual to self helicity of two interacting loops, and De~Pontieu et al. (\cite{depo14b}) found the presence of twisting motions to be ubiquitous.

These helical twisting motions might be related to the red-blue-asymmetries. The asymmetry is changing along the loop, i.e., while we follow the ejected plasma packet (Figs.\,\ref{F:statistics}c and d). Therefore the twisting motions could show themselves as enhancements in the line wings. Should this be the case, we would expect an oscillatory variation of the line asymmetry, i.e., with the twisting moving back and forth we would see the enhancements alternating between the red and blue wings. Instead, after the initial phase we see only a small asymmetry which is mostly slightly leaning to the blue wing (see Fig.\,\ref{F:statistics}). Thus, we do not think that the line asymmetries are caused by helical twisting motions. Other possibilities for a (symmetric) enhancement of the line wings might be a change of the velocity distribution of the ions due to ion heating by Alfv\'en waves or due to the motions directly associated with the different modes of MHD waves (Peter~\cite{peter10}; his Sects. 4.3 and 4.5.1).

The presence of a helical structure within the loop hints at the driving of the loop through footpoint motions. The original idea of Parker (\cite{parker72,parker88}) was that irreversible horizontal motions in the photosphere braid the magnetic field and induce currents in the corona that will be dissipated in a transient fashion in the form of nanoflares. Numerous models have been constructed that use (simple helical) motions at the loop footpoints and investigated how these motions drive the heating and dynamics in the coronal part of a loop (e.g.,\ Wilmot-Smith et al. \cite{wilmot11}). Alternatively, some studies assumed a braided state of the loop as an initial condition (that would have been induced by e.g., helical motions) and concentrated on the relaxation phase when the magnetic field reconnects and heats the plasma (e.g., Pontin et al. \cite{pontin17}).  Models with sufficient resolution find the magnetic field in the loop to be in  a turbulent state (e.g., Reid \cite{reid18}, their Fig.\,5). If the magnetic field is in a turbulent state, also the motions along the magnetic field will be distributed in all directions (and not only along the major axis, viz.,\ the guide field, of the loop). Consequently we can expect significant line broadening in this situation, which might show up as a broadening of the line core and as a broad additional component in the emission line profiles, i.e.,\ as enhanced wings. In our observations we see both the increased line width in the loop (cf.\ Fig.\,\ref{F:loop}d) and excess emission in both line wings (e.g.,\ Fig.\,\ref{F:lineprofiles}, x11 to x25).

If turbulence is present in the loop, it would also go along with heating of the plasma and consequently an increase in the emission from the loop (Fig.\,\ref{F:statistics}a). The large line width in the loop (Fig.\,\ref{F:loop}d) also supports that the plasma is heated. Considering the contribution function of \ion{Si}{iv}, only a modest increase of the temperature from, e.g., 60\,000\,K to below 80\,000\,K would be sufficient to explain an increase in intensity by a factor of two, consistent with Fig.\,\ref{F:statistics}a.

In general, in these 3D loop models helical motions will lead to a braided state that in turn results in a turbulent state of the magnetic field and conversion from magnetic to thermal (and kinetic) energy. Future numerical loop models will have to investigate if the turbulent state of the magnetic field will indeed result in a line broadening and wing enhancement as we observe here. Only then we can clearly link the presence of helical motions in the loop to the enhanced emission in the line wings.

\subsection{Presence of a coronal component in the cool loop}\label{S:coronal.comp}

%
If heating is present in the loop, (parts of) the plasma might be heated also to coronal temperatures. Our investigation so far concentrated on the spectral profiles of the \ion{Si}{iv} line. In equilibrium, this line forms below 0.1\,MK and in the bright parts of an active region
it will dominate the slit-jaw images of IRIS in the 1400\,{\AA} band. We see the loop under investigation also in AIA data, e.g. in the 171\,{\AA} band (Fig.\,\ref{F:loop}f). The loop is visible also in other AIA channels (e.g.,\ 304\,{\AA}, 131\,{\AA}, and 193\,{\AA}), but we will concentrate on the 171\,{\AA} band here.

That we see emission at 171\,{\AA} is not sufficient to argue that we indeed see hot coronal plasma here. Usually, the bulk of the 171\,{\AA} band is dominated by \ion{Fe}{ix} forming at coronal temperatures just below 1\,MK. But there is also a contribution from lower temperatures around 0.2\,MK to 0.3\,MK (e.g.,\ Boerner et al.\ \cite{boerner12}).  Because we see the loop  in  \ion{O}{iv} 1401\,{\AA} forming at about 0.2\,MK, one might wonder if the loop at  171\,{\AA} might be due to cool plasma at (a few) 0.1\,MK and does not represent coronal emission at about 1\,MK.

\begin{figure}
   \centering
   \includegraphics[width=88mm]{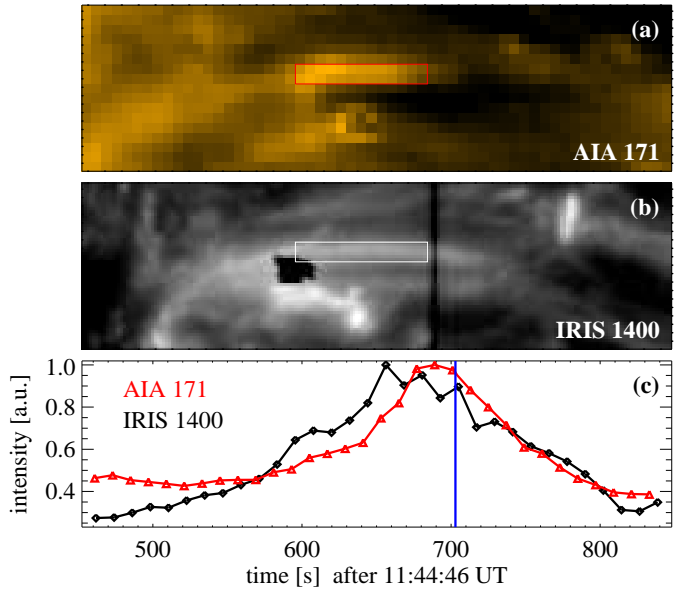}
   \caption{Temporal evolution of loop intensity.
   Panels (a) and (b) show snapshots in the AIA\,171\,{\AA} channel and the
IRIS slit-jaw images at 1400\,{\AA}. The field of view is the same as in
Fig.\,\ref{F:loop}, but here we use a logarithmic  scaling of the intensities. The apex region of the loop under investigation here is indicated
by the boxes in panels (a) and (b). Panel (c) displays the intensity integrated
over the box in the two channels. The vertical blue line shows
the time of the snapshots in panels (a) and (b). The black vertical line
in panel (b) is the spectrograph slit, the black feature at the bottom left
edge of the box in panel (b) is an artifact on the detector. An animation
showing the temporal evolution while the slit is passing the loop is available
online. See Sect.\,\ref{S:coronal.comp}.}

    \label{F:coronal.comp}%
\end{figure}

To test this, we compare the temporal evolution of the images in AIA 171\,{\AA} to  the IRIS slit-jaw images at 1400\,{\AA}. If 171\,{\AA} would be dominated by cool (few 0.1\,MK) plasma, then 171\,{\AA} and 1400\,{\AA} should show the same temporal evolution. In Fig.\,\ref{F:coronal.comp} we show snapshots in the two bands together with the evolution of the emission from the apex region of the loop. The movie that is attached to Fig.\,\ref{F:coronal.comp} displays the evolution in 171\,{\AA} and 1400\,{\AA} over the time the slit crosses the loop. There we  see the apparent motion along the loop in 1400\,{\AA} that is captured in the space-time plot in Fig.\,\ref{F:propermotion}. We also see a brightening near the loop apex in 171\,{\AA}, but because of the significantly lower spatial resolution of AIA compared to IRIS, apparent motions are more difficult to isolate on these scales. Still, we can study the temporal variation in the region around the apex of the flat loop that we here approximate by a rectangle (Figs.\,\ref{F:coronal.comp}a,b). We integrate the emission in 171\,{\AA} and 1400\,{\AA} in this rectangle and show the resulting light curves in Fig.\,\ref{F:coronal.comp}c. There is a clear time lag between the two channels, with the 171\,{\AA} channel brightening about half a minute to one minute after the 1400\,{\AA} channel. This time lag shows that 171\,{\AA} in this loop can not be dominated by cool pülasma. Hence the loop we see in 171\,{\AA} should originate from coronal plasma.
%

Instead, the time lag between the 1400\,{\AA} channel (0.1\,MK) and the 171\,{\AA} band (1\,MK) indicates that the plasma is heated. Thus we find further supporting evidence for heating in the loop as discussed in Sect.\,\ref{S:disc.helical}.
%

\section{Conclusions}

Using the data of an emerging active region mapped by IRIS, we studied the evolution of plasma injected into a cool transition region loop by investigating the details of the  \ion{Si}{iv} (1394\,\AA) line profiles. In this low-lying loop, at some 12\,Mm in length it climbs probably less than 1\,Mm, a steady  flow from the eastern to the western side is established (Sect.\,\ref{S:disc.geometry}).

At the eastern footpoint we see a clear high-speed injection of plasma into the loop, revealed by a strong excess of the emission in the blue wing of the \ion{Si}{iv} line profile (Sect.\,\ref{S:disc.heating}). At the same time, the line core is redshifted at this same footpoint which is indicative of a strong heating event at temperatures just above where \ion{Si}{iv} forms. This would cause a (local) pressure enhancement that would  press the bulk of the plasma down, while some part is injected into the loop.

In the loop we find a signature for a helical flow which  implies that also the magnetic field is helical (Sect.\,\ref{S:disc.helical}). This could be a signature of a driving of the loop by circular horizontal motions at its footpoint in the photosphere. This could lead not only to twisting motions but also to a turbulent state of the magnetic field. The associated (small-scale non-resolved) flows could then be responsible for the (almost) symmetric enhancement of both wings of the line profile in the middle section of the loop away from its footpoints  (Sect.\,\ref{S:disc.helical}). Such profiles with symmetrically enhanced wings have been observed before (see Sect.\,\ref{S:intro} and e.g.,\ Peter \cite{peter10}), but so far await a solid interpretation. With new models of MHD turbulence in a loop we might get a better understanding of the nature of these profiles.

\begin{acknowledgements}

Sincere thanks are due to Davina Innes and Lijia Guo for discussions and comments on this study.
HP is grateful for discussions on the turbulent state of the magnetic field with Pradeep Chitta and David Pontin. 
The authors thank the anonymous referee for helpful comments.
IRIS is a NASA small explorer mission developed and operated by LMSAL with mission operations executed at NASA Ames Research center and major contributions to downlink communications funded by ESA and the Norwegian Space Centre.
The AIA and HMI data used are provided courtesy of NASA/SDO and the
AIA and HMI science teams.
This work is supported by the National Foundations of China (11673034, 11533008, 11790304 and 11773039) and Key Programs of the Chinese Academy of Sciences (QYZDJ-SSW-SLH050).

\end{acknowledgements}

\clearpage

\begin{appendix} 

\section{The profiles of \ion{Si}{iv} along the loop in the center of the loop and at its edges\label{appendix}}

In our study we concentrate on the central spine of the loop (viz.,\ the central axis of the loop, see dotted line in Fig.\,\ref{F:loop}b). Here we show that those results are not just valid on the thin on-axis part of the loop. We do this by checking the line profiles and profile parameters not just at the central loop axis, but also at spatial positions further away. Because the loop runs roughly in the east-west direction, we here investigate the region up to two spatial IRIS pixels (0.33\arcsec\ or 250\,km) to the north and south of the loop center. The loop has a width of about 0.6\,Mm (full width at half maximum, see Sect.\,\ref{S:overall}), so these locations are still within the loop, but at its edge.

\begin{figure*}
\centering
\includegraphics[width=180mm]{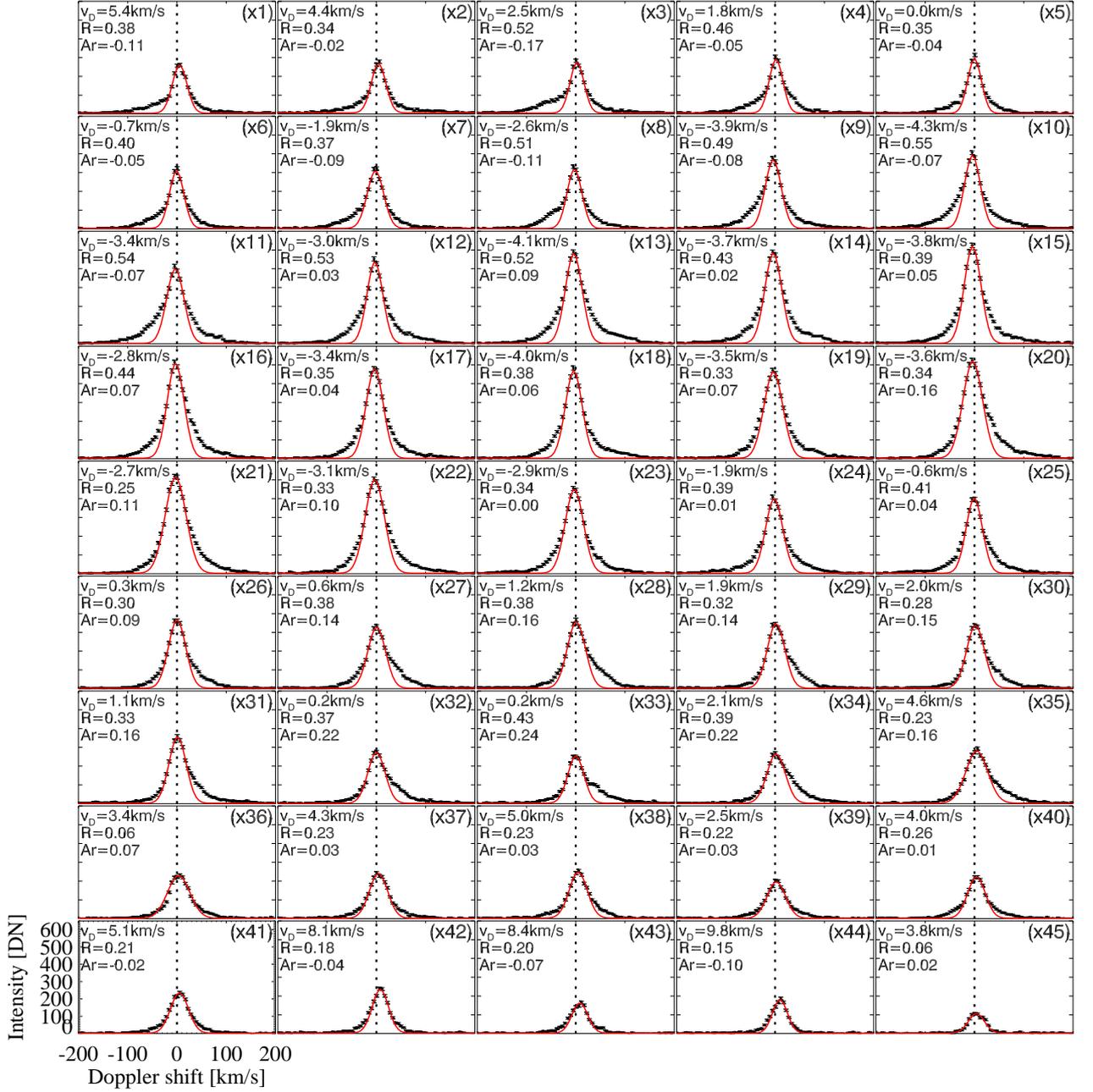}
\caption{Same as Fig.\,\ref{F:lineprofiles}, but for the positions
two spatial pixels below (south of) the pixels where the \ion{Si}{iv} (1394\,\AA) spectra in  Fig.\,\ref{F:lineprofiles} are taken. This essentially follows the southern edge of the loop. The central axis of the loop is shown in Fig.\,\ref{F:loop}b.
The diamonds (with the bars) show the observed spectra
and the red lines indicate single-Gaussian fits
to the line core.   The parameters denoted by the numbers in the plots
show line centroid ($v_{D}$) of the red fits, the total residual of
the intensity ($R$), and the red-blue asymmetry ($A_{\rm{r}}$).}
\label{lineprofiles_m}
\end{figure*}

\begin{figure*}
\centering
\includegraphics[width=180mm]{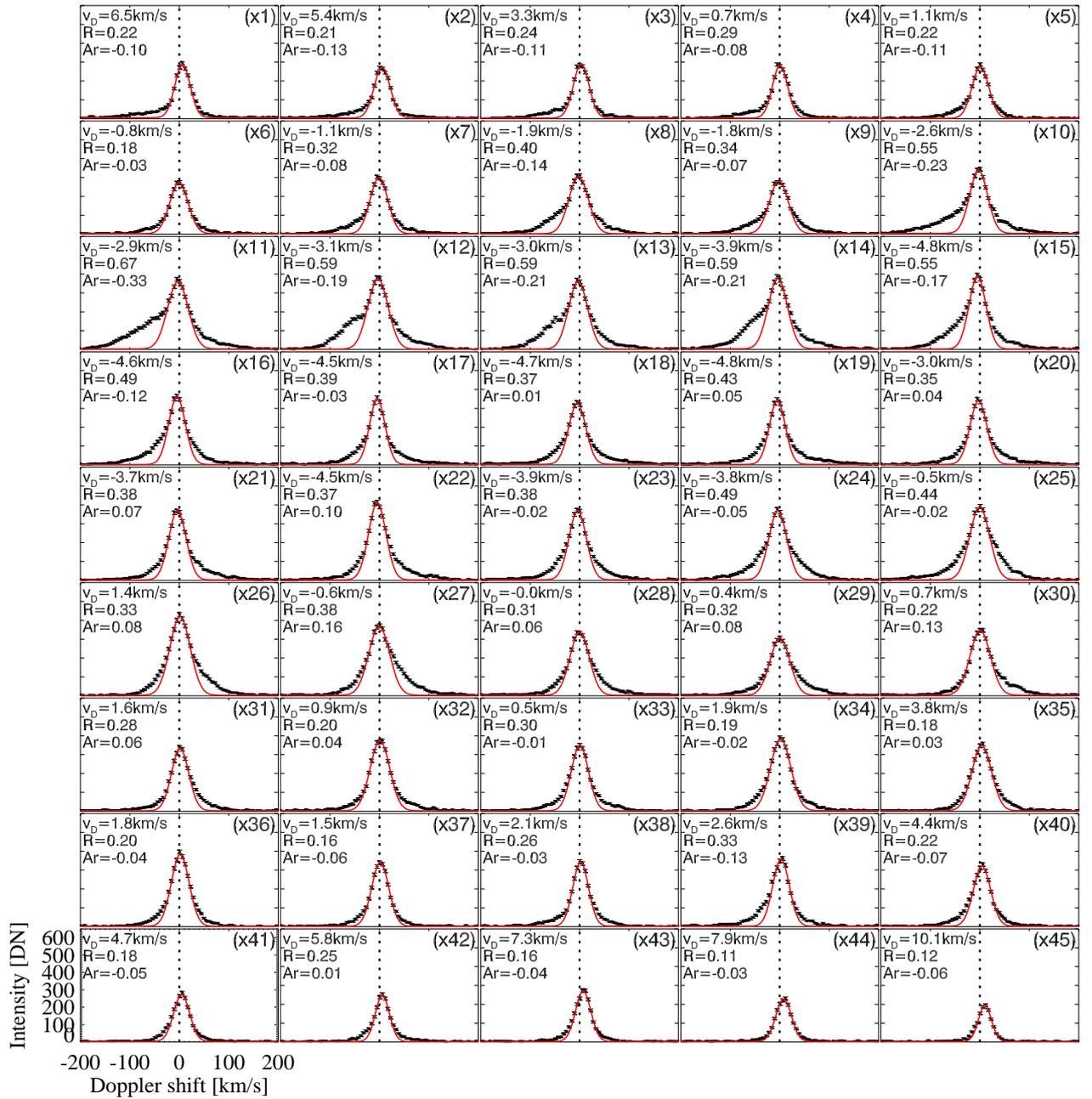}
\caption{Same as Fig.\,\ref{lineprofiles_m}, but now for the positions
two spatial pixels above (north of) the pixels where the \ion{Si}{iv} (1394\,\AA)
spectra in  Fig.\,\ref{F:lineprofiles} are taken.}
\label{lineprofiles_p}
\end{figure*}

To check the shape of the line profiles, we show the \ion{Si}{iv} line two pixels below (south; Fig.\,\ref{lineprofiles_m}) and above (north; Fig.\,\ref{lineprofiles_p}) the spatial positions that we used in Fig.\,\ref{F:lineprofiles}. Otherwise Figs.\,\ref{lineprofiles_m} and \ref{lineprofiles_p} are exactly the same as Fig.\,\ref{F:lineprofiles}. The prominent enhanced emission in the blue wing near the eastern footpoint, the nearly single-Gaussian profiles near the western footpoint, and the evolution of the enhancements in both line wings along the loop are overall very similar at the edge of the loop and at its center. Therefore, the discussion in Sect.\,\ref{S:spectra} does apply to the whole loop and not only its central axis.

\begin{figure}
\centering
\includegraphics[width=88mm]{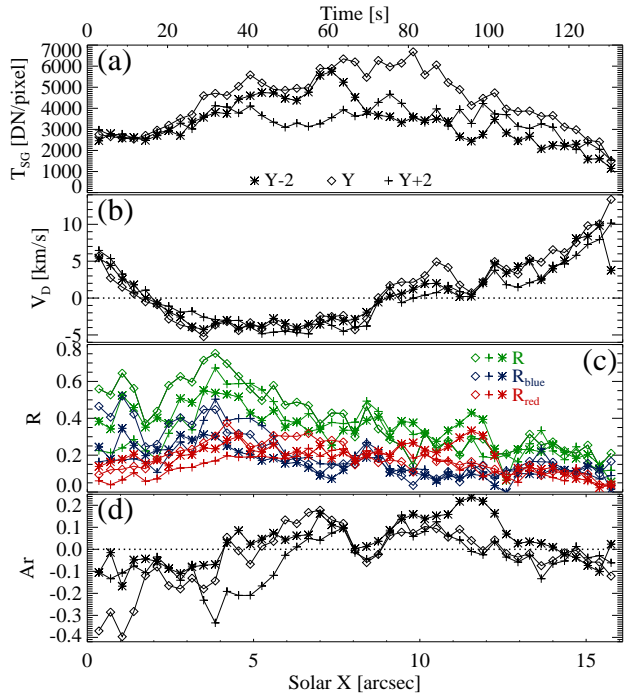}
\caption{Same as Fig.\,\ref{F:statistics}, but for the positions
two spatial pixels below (south of) and above (north of) the pixels where the  line profile parameters in Fig.\,\ref{F:statistics} are taken. This essentially follows
the southern and northern edges of the loop. The variation along the southern edge of the loop is denoted by asterisks, the variation along the northern edge by crosses. The diamonds show the variation along the central loop axis (cf. dotted line in Fig.\,\ref{F:loop}b) and are identical to the data shown in Fig.\,\ref{F:statistics}.}
\label{statistics_mp}
\end{figure}

\begin{figure}
\centering
\includegraphics[width=88mm]{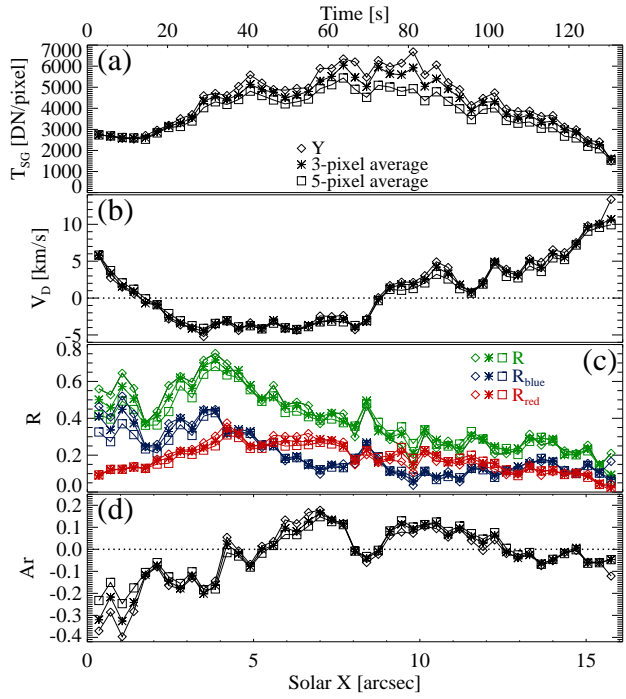}
\caption{Similar to Fig.\,\ref{statistics_mp}, but for the parameters derived from profiles averaged in the north-south direction across the loop.  Squares denote averages over five spatial pixels, asterisks show the averages over three pixels. The diamonds show the variation along the central loop axis and are identical to the data shown in Fig.\,\ref{F:statistics}.} \label{statistics_av}
\end{figure}

Also, the loop-profile parameters as discussed in Sect.\,\ref{S:evolution.along.loop} and shown in Fig.\,\ref{F:statistics} are taken at the central axis of the loop. To check if our results  apply to the whole part of the loop, too, we check them north and south of the loop center (Fig.\,\ref{statistics_mp}) and look at averages across the loop (Fig.\,\ref{statistics_av}). We first calculate the line profile parameters at two pixels below and above the loop, i.e.,\ for the profiles shown in Figs.\,\ref{lineprofiles_m} and \ref{lineprofiles_p}. The results are shown in Fig.\,\ref{statistics_mp} together with the parameters for the center of the loop as they are displayed in Fig.\,\ref{F:statistics}. The variation of the line profile parameters along the loop is slightly different, but overall the quantitative variation is very similar. As a second test, we average the line profiles in the north-south direction centered around the central loop axis and calculate the parameters from those averaged profiles. We plot the results in Fig.\,\ref{statistics_av}, again together with the results for the profiles along the central axis. The parameters of the averaged profiles match very well with those at the central loop axis.

This underlines that the results for the central axis of the loop are robust. The similarity of the line profiles and the good match of the variation of the line profile parameters along the loop shows that in the framework of this study we can consider this loop as a more or less monolithic structure.

\end{appendix}


\clearpage

\end{document}